\documentclass[2p,sort&compress]{elsarticle}
\usepackage{lineno,hyperref}
\usepackage{amsmath,bm,accents}
\usepackage{amssymb}

\usepackage{mathrsfs}

\modulolinenumbers[5]

\makeatletter
\def\ps@pprintTitle{%
 \let\@oddhead\@empty
 \let\@evenhead\@empty
 \def\@oddfoot{}%
 \let\@evenfoot\@oddfoot}
\makeatother
\journal{}

\begin{document}

\begin{frontmatter}

\title{{\bf A thermodynamically consistent formulation of generalized thermoelasticity at finite deformations}}

\author[a,b]{M. F. Wakeni\corref{mycorrespondingauthor}}
\cortext[mycorrespondingauthor]{Corresponding author}
\ead{wknmeb001@myuct.ac.za}

\author[a,b]{B.D. Reddy}
\author[a]{A.T. McBride}
\address[a]{Centre for Research in Computational and Applied Mechanics (CERECAM),
University of Cape Town, 7701 Rondebosch, South Africa}
\address[b]{Department of Mathematics and Applied Mathematics,
University of Cape Town, 7701 Rondebosch, South Africa}

\begin{abstract}
A thermodynamically consistent model of non-classical coupled non-linear thermoelasticity capable of accounting for thermal wave propagation is proposed. The heat flux is assumed to consist of both additive energetic and dissipative components. Constitutive relations for the stress, the entropy and the energetic component of the heat flux are derived in a thermodynamically consistent manner. A Lyapunov function for the dynamics is obtained for the case in which the surface of the continuum body is maintained at a reference temperature. It is shown that the system is non-linearly stable. The linearized model is shown to be similar to the type III model of Green and Naghdi, except for some minor differences in the interpretations of some of the parameters.
\end{abstract}

\begin{keyword}
Thermodynamically consistent; Thermoelasticity; Energetic and dissipative flux component; Constitutive relations; Lyapunov function; Type III model 
\end{keyword}

\end{frontmatter}
\section{Introduction}
The propagation of heat energy in a rigid body is governed by the energy balance equation which reads
\begin{equation}\label{eq:en-bal}
\dot{e} = -\mathrm{div}\,\mathbf{q} + r.
\end{equation} 
Here $e$, $\mathbf{q}$, and $r$ denote the internal energy, heat flux, and heat source respectively. The superposed dot denotes time derivative and $\mathrm{div}[\,\cdot\,]$ divergence. In the classical theory of heat conduction with Fourier's law \cite{Fourier1822}, the balance equation \eqref{eq:en-bal} is supplemented by a constitutive equation for the heat flux $\mathbf{q}$ satisfying the inequality
\begin{equation}\label{eq:heat-cond}
\mathbf{q}\cdot\nabla\Theta \leq 0,
\end{equation} 
where $\Theta$ is the absolute temperature. The inequality \eqref{eq:heat-cond}, also referred to as \emph{heat conduction inequality}, states that heat only flows by the mechanism of diffusion from regions of higher to those of lower temperature distributions. In other words, the vector quantity describing the heat flow, that is, the heat flux $\mathbf{q}$, always points in the direction of decreasing temperature distribution. 

However, for example, in the linear case this mechanism leads to a parabolic partial differential equation which is characterised by infinite speed of propagation of localized thermal disturbances, a paradoxical phenomenon from a physical point of view.  

Despite this non-physical prediction, the classical theory of heat conduction by Fourier's law has been successful for a broad range of engineering applications. However, as length scales decrease, this phenomenon of instantaneous propagation of thermal disturbances becomes more dominant at low ranges of temperature near absolute zero, so that approximations used in the classical theory of heat conduction lose validity \cite{Fryer2013a}. 

The earliest known conjecture on the existence of thermal propagation as waves, also known as the second sound phenomenon, was given by Nernst \cite{Nernst1918} in 1917. Later, in 1938 Tisze \cite{Tisza1938} and  Landau \cite{Landau1941} in 1941 independently suggested the possibility of thermal waves in superfluid liquid helium, at temperatures below the so-called lambda transition near 2.2 K \cite{Donnelly2009a}. 
Peshkov \cite{peshkov2013second} reported the first experimental evidence for the existence of second sound in 2He. In his work, Peshkov suggests that second sound might also be observed in pure crystalline materials on the basis of similarities between crystal materials and liquid helium. Laser pulsing experiments have shown that second sound can propagate in high-purity crystals of 4He \cite{ackerman1966second}, 3He⁠ \cite{ackerman1969second}⁠,  NaF\cite{jackson1970second} , and Bi \cite{narayanamurti1972observation}. 

Cattaneo \cite{Cattaneo1958} was the first to introduce a non-Fourier theory of heat conduction in order to overcome the paradoxical prediction of the classical Fourier theory. His work is based on the concept of relaxing the heat flux from the classical Fourier law to obtain a constitutive relation with a non-Fourier effect. There have been several other attempts to develop continuum theories capable of predicting thermal waves propagating at finite speeds for various types of media. Among these, the works in \cite{Lord1967, Gurtin1968, fox1969generalised, Muller1971, Green1972} are notable. The constitutive equation for the heat flux according to Cattaneo is examined in terms of thermodynamics in \cite{Coleman1982, Coleman1982b}. Later this result was extended in \cite{Oncu1992} to the case where deformation is applicable.

 A relatively more recent theory of non-classical heat conduction with and without deformation was proposed by Green and Naghdi  \cite{Green1991, Green1992, Green1993, Green1995}. Their work is based on the introduction of three types of constitutive relation for the heat flux, thereby resulting in three different models, namely type I, which is the classical theory, type II, a purely hyperbolic model which allows the propagation of a heat pulse without damping, and type III, which is the combination of the first two. 

In recent years there has been a considerable amount of interest concerning the theory of Green and Naghdi. An extensive overview of the theory can be found in \cite{Chandra1996, Chandra1998, Hetnarski1999}. Theoretical results addressing existence and uniqueness \cite{Quintanilla2002, Quintanilla2002a} and  exponential stability \cite{Racke2002} have also been investigated for some types of the theory. The designing of appropriate numerical methods has also been addressed in \cite{Bargmann2008b,Wakeni2015}.

This work is concerned with a thermodynamically consistent formulation of a fully non-linear coupled problem of non-classical thermoelasticity inspired by that of Green and Naghdi. The formulation is based on the basic laws of continuum thermodynamics, the balance laws of momentum, the balance of energy, and the entropy imbalance. However, the point of departure from the classical theory comes from two assumptions: the first is that the heat flux is additively composed of two parts, namely the dissipative and energetic components, and the second is that a material derivative of a time primitive of the absolute temperature is assumed to be the proportionality constant of the heat and the entropy conjugate pairs.

 Thermodynamic restrictions on the constitutive relations are derived using the procedure of Coleman and Noll \cite{Coleman1963}. Stability of the system of partial differential equations governing the thermomechanical coupling in the non-classical regime is proved in the sense of Lyapunov. The other notable aspect of the model is that the linearized theory is similar to that of by Green and Naghdi except for some differences in the interpretation of the material parameters.

The rest of the paper is organised as follows. In Section \ref{sec:kinematics}, geometric and kinematical descriptions of the continuum body are presented. In Section \ref{sec:thermo-mechanics} the non-classical model describing the coupling of mechanical deformation and non-classical heat conduction is formulated based on the laws of thermodynamics. Constitutive relations for the stress, the entropy and the energetic component of the heat flux are derived from a free energy via the Coleman-Noll procedure. Next, in Section \ref{sec:IBVP} the initial boundary value problem (IBVP) of non-classical thermoelasticity is summarised. A class of physically meaningful initial and boundary conditions are also proposed. A Lyapunov function for the dynamics generated by the IBVP rendering the system  non-linearly stable is obtained in Section \ref{sec:stability}. Finally in Section \ref{sec:linear-theory}, the linearised form of the IBVP is summarised.

\section{Kinematics: Lagrangian description}\label{sec:kinematics}
Consider a continuum body occupying the reference configuration $\Omega\subset\mathbb{R}^{d}$, where $d=1$, $2,$ or $3$, with smooth boundary $\Gamma$ and material point denoted by $\mathbf{X}\in\Omega$. The surface outward unit normal in the reference configuration is denoted by $\mathbf{N}$. The motion of the body is parametrized by a $C^1$ orientation-preserving map:
\begin{equation}
\bm{\varphi}(\mathbf{X},t):\mathbb{I}\to\mathbb{R}^{d},\quad\quad \mathbf{X}\in\overline{\Omega},
\end{equation} 
where $\overline{\Omega}$ represents the closure of $\Omega$, or the union of the interior of $\Omega$ and its boundary $\Gamma$, and $\mathbb{I}=[0,T]$ denotes the time domain. The displacement and velocity fields in the \emph{Lagrangian description} associated with a particle located at point $\mathbf{X}$ at time $t\in\mathbb{I}$ respectively are
\begin{equation}
\bm{u}(\mathbf{X}, t) = \bm{\varphi}(\mathbf{X}, t) - \mathbf{X}, \qquad\bm{v}(\mathbf{X}, t) = \dfrac{\partial}{\partial t}~\bm{u}(\mathbf{X}, t). 
\end{equation}
In what follows, partial time derivatives will be denoted by superimposed dots so that $\dot{\bm{\diamond}}={\partial \bm{\diamond}}/\partial t$ for some field $\bm{\diamond}$ in the material description. The deformation gradient $\mathbf{F}$, the primary measure of deformation, is defined by
\begin{equation}
\mathbf{F}=\nabla\bm{u}+\mathbf{1},
\end{equation}
where the operator $\nabla$ denotes the gradient with respect to the reference configuration, and $\mathbf{1}$ is the identity second-order tensor. For an arbitrary material point $\mathbf{X}$ and time $t\in \mathbb{I}$, an infinitesimal reference volume element $\mathrm{d}\Omega$ associated with the material point is mapped into a volume element $\mathrm{d}\omega$ in the current configuration by the Jacobian $J(\bm{X}, t)=\mathrm{det}\mathbf{F}(\bm{X}, t)$ via the relation
\begin{equation}
\mathrm{d}\omega=J(\bm{X}, t)\mathrm{d}\Omega.
\end{equation} 
Thus, an admissible configuration is one in which the Jacobian $J$ is positive. 
\section{Non-linear coupled thermo-mechanics in the non-classical framework}\label{sec:thermo-mechanics}
%
%
\subsection{Balance laws}
\begin{itemize}
\item The local forms of the momentum balance are 
\begin{equation}\label{eq:Lmomu-bal}
\rho_0\ddot{\bm{u}} = \mathrm{Div}\mathbf{P} +\rho_0\bm{b},
\end{equation}
\begin{equation}\label{eq:Amomu-bal}
\mathbf{P}\mathbf{F}^{\texttt{T}} =\mathbf{F}\mathbf{P}^{\texttt{T}}.
\end{equation}
Here $\rho_0$, $\bm{b}$ and $\mathbf{P}$  are, respectively, the density in the reference configuration, the body force density and the first Piola-Kirchhoff stress tensor. The superscript $\texttt{T}$ denotes the transpose operation on second-order tensors.
\item The balance of energy (first law of thermodynamics) is given by
\begin{equation}\label{eq:Energy-bal}
\dot{\mathcal{E}}=\mathbf{P}:\dot{\mathbf{F}}-\mathrm{Div}\mathbf{Q}+ R,
\end{equation}
where $\mathcal{E}$, $\mathbf{Q}$, and $R$ are the internal energy, heat flux, and heat source.
\item The entropy imbalance (second law of thermodynamics) is given by the inequality
\begin{equation}\label{eq:en-imbalance}
\dot{\eta}\geq-\mathrm{Div}\mathbf{H}+S,
\end{equation}
where $\eta$, $\mathbf{H}$, and $S$ are the entropy, entropy flux vector, and entropy source, respectively.
\end{itemize}
The absolute temperature $\Theta$ is used to relate heat flux (resp. heat source) with entropy flux (resp. entropy source) in the classical theory (see, for example \cite{Gurtin2010} pp. 187). In our formulation, motivated by the theory of Green and Naghdi, we additionally assume that the existence of a scalar field $\alpha$, referred to as the \emph{thermal displacement}, such that
\begin{equation}
\dot{\alpha}> 0,
\end{equation}
and that 
\begin{equation}\label{eq:entropy-heat}
\mathbf{H}=\dfrac{\mathbf{Q}}{\dot{\alpha}},\quad \text{ and }\quad S = \dfrac{R}{\dot{\alpha}}.
\end{equation} 
We then proceed to define the absolute temperature by 
\begin{equation*}
\Theta = \dot{\alpha}.
\end{equation*} 

Next, substitution of \eqref{eq:entropy-heat} into \eqref{eq:en-imbalance} leads to 
 \begin{equation}\label{eq:en-imbalance2}
\dot{\eta}\geq-\mathrm{Div}\bigg[\dfrac{\mathbf{Q}}{\Theta}\bigg]+\dfrac{R}{\Theta}\quad \text{ or }\quad \Theta\dot{\eta}\geq-\mathrm{Div}\mathbf{Q}+\dfrac{1}{\Theta}\mathbf{Q}\cdot\nabla\Theta + R.
\end{equation}
By subtracting the inequality \eqref{eq:en-imbalance2} from the energy balance equation \eqref{eq:Energy-bal} and subtracting $\dot{\Theta}\eta$ from both sides of the resulting inequality, we obtain
\begin{equation}\label{eq:enter-imbal}
\dfrac{\partial}{\partial t}\big[\mathcal{E}-\Theta\eta\big]\leq \mathbf{P}:\dot{\mathbf{F}}-\dot{\Theta}\eta-\dfrac{1}{\Theta}\mathbf{Q}\cdot\nabla\Theta.
\end{equation}
On defining the Helmholtz free-energy through the Legendre transformation
\begin{equation}\label{eq:legendre}
\Psi = \mathcal{E}-\Theta\eta,
\end{equation}
\eqref{eq:enter-imbal} yields the local form of the free-energy imbalance
\begin{equation}\label{eq:free-en-imbal}
\dot{\Psi} - \mathbf{P}:\dot{\mathbf{F}}+\dot{\Theta}\eta+\dfrac{1}{\Theta}\mathbf{Q}\cdot\nabla\Theta  = -\Theta\Xi\leq 0.
\end{equation}
Here $\Xi$ denotes the rate of entropy production (while $\Theta\Xi$ is the rate of energy dissipation).
\subsection{Constitutive theory}
As mentioned above in the motivation, we assume that the heat flux $\mathbf{Q}$ is split additively as
\begin{equation}\label{eq:heat-flux}
\mathbf{Q}=\mathbf{Q}_E+\mathbf{Q}_D.
\end{equation}
We refer to $\mathbf{Q}_E$ and $\mathbf{Q}_D$ as the \emph{energetic} and \emph{dissipative} components of the heat flux $\mathbf{Q}$. As a consequence of this split, the free-energy imbalance \eqref{eq:free-en-imbal} becomes \begin{equation}\label{eq:free-en-imbal2}
\dot{\Psi} - \mathbf{P}:\dot{\mathbf{F}}+\dot{\Theta}\eta+\dfrac{1}{\Theta}\mathbf{Q}_E\cdot\nabla\Theta+ \dot{\Theta}\eta+\dfrac{1}{\Theta}\mathbf{Q}_D\cdot\nabla\Theta = -\Theta\Xi\leq 0.
\end{equation} 

Looking at the variables involved in the free-energy imbalance \eqref{eq:free-en-imbal2}, we assume that the state of the system under consideration depends on the set of state variables
\begin{equation}\label{eq:state-vars}
\mathscr{A} = \left\lbrace \mathbf{F},\Theta, \nabla\alpha, \nabla\Theta\right\rbrace.
\end{equation}
Thus the free energy $\Psi$, the Piola stress $\mathbf{P}$, the entropy $\eta$, and the heat flux components $\mathbf{Q}_E$ and $\mathbf{Q}_D$ are determined by constitutive equations of the form

\begin{xalignat}{6}\label{eq:constitutive}
\nonumber & &\Psi~ &=~\Psi(\mathscr{A}), & \mathbf{P}~ &=~\mathbf{P}(\mathscr{A}),& \eta &= \eta(\mathscr{A}), & &\\
& &\mathbf{Q}_E &=\mathbf{Q}_E(\mathscr{A}), & \mathbf{Q}_D &=\mathbf{Q}_D(\mathscr{A}).& & &
\end{xalignat}
We now apply the Coleman--Noll procedure to ensure that the constitutive equations \eqref{eq:constitutive} satisfy the laws of thermodynamics. If the constitutive equation for the free energy is differentiated with respect to time, one obtains
\begin{equation}\label{eq:dfree}
\dot{\Psi}=\dfrac{\partial \Psi}{\partial \mathbf{F}}:\dot{\mathbf{F}}+ \dfrac{\partial \Psi}{\partial \Theta}\;\dot{\Theta}+\dfrac{\partial \Psi}{\partial \bm{\Lambda}}\cdot\dot{\bm{\Lambda}} +\dfrac{\partial \Psi}{\partial \mathbf{G}}\cdot\dot{\mathbf{G}},
\end{equation}
where $\bm{\Lambda}=\nabla\alpha$ and $\mathbf{G}=\nabla\Theta$.

Substitution of the time derivative of the free energy \eqref{eq:dfree} into the free imbalance \eqref{eq:free-en-imbal2} yields
\begin{equation}\label{eq:Col-Nol}
\bigg(\dfrac{\partial \Psi}{\partial \mathbf{F}}-\mathbf{P}\bigg):\dot{\mathbf{F}}+
\bigg(\dfrac{\partial \Psi}{\partial \Theta}+\eta\bigg)\;\dot{\Theta}+
\bigg(\dfrac{\partial \Psi}{\partial \bm{\Lambda}}+\dfrac{1}{\Theta}\mathbf{Q}_E\bigg)\cdot\dot{\bm{\Lambda}}+
\dfrac{\partial \Psi}{\partial \mathbf{G}}\cdot\dot{\mathbf{G}}+
\dfrac{1}{\Theta}\mathbf{Q}_D\cdot\nabla\Theta\leq 0,
\end{equation}
which must be satisfied for all states.

Since $\dot{\mathbf{F}}$, $\dot{\Theta}$, $\dot{\bm{\Lambda}}$, $\dot{\mathbf{G}}$, and their time primitives can be chosen arbitrarily in order to maintain the inequality \eqref{eq:Col-Nol} we may choose the constitutive equations and thermodynamic restrictions as
\begin{equation}\label{eq:const-rels}
\mathbf{P} = \dfrac{\partial \Psi}{\partial \mathbf{F}}, \quad \eta = -\dfrac{\partial \Psi}{\partial \Theta}, \quad \dfrac{1}{\Theta}\mathbf{Q}_E = -\dfrac{\partial \Psi}{\partial \bm{\Lambda}},\quad\dfrac{\partial \Psi}{\partial \mathbf{G}} = \mathbf{0}, \quad \text{and }~\mathbf{Q}_D\cdot\nabla\Theta\leq 0. 
\end{equation}
As consequence, the rate of entropy production in any thermodynamically admissible process becomes
\begin{equation*}
\Xi = -\dfrac{1}{\Theta^2}\mathbf{Q}_D\cdot\nabla\Theta.
\end{equation*}
Note that $\dot{\bm{\Lambda}}=\nabla\Theta$.
Equation \eqref{eq:const-rels}$_{4}$ reveals that the free energy does not depend on $\nabla\Theta$.

For instance, a class of such free energy functions in the non-classical case has the form
\begin{equation}\label{eq:ex-free-enr}
\Psi=\Psi_{\mathrm{c}}+\dfrac{1}{2}\mathbf{K}_1\nabla\alpha\cdot\nabla\alpha,
\end{equation}
where $\Psi_c$ denotes any classical free energy function for thermo-hyperelasticity  and $\mathbf{K}_1$ a symmetric and positive-definite non-classical  conductivity second-order tensor. Equation \eqref{eq:ex-free-enr} and the constitutive equation \eqref{eq:const-rels}$_3$ imply that the energetic component of heat flux $\mathbf{Q}_E$ is linear in $\nabla\alpha$. 
\subsection{Field equation for the thermal conduction}
 Consequently, by \eqref{eq:dfree} and \eqref{eq:const-rels}, the time derivative of the free energy becomes
\begin{equation}\label{eq:dfree2}
\dot{\Psi}=\mathbf{P}:\dot{\mathbf{F}}-\eta\dot{\Theta}-\dfrac{1}{\Theta}\mathbf{Q}_E\cdot\dot{\bm{\Lambda}}.
\end{equation}

Next, by the Legendre transformation \eqref{eq:legendre}, from \eqref{eq:dfree2} the time derivative of the internal energy is
\begin{equation}\label{eq:dinternal}
\dot{\mathcal{E}}= \Theta\dot{\eta}+\mathbf{P}:\dot{\mathbf{F}}-\dfrac{1}{\Theta}\hat{\mathbf{Q}}_E\cdot\dot{\bm{\Lambda}}.
\end{equation}
By combining \eqref{eq:Energy-bal} and \eqref{eq:dinternal} we obtain the field equation of thermal conduction in the conservation form 
\begin{equation}\label{eq:cons-form}
\Theta\dot{\eta}=-\mathrm{Div}\mathbf{Q} +\dfrac{1}{\Theta}\mathbf{Q}_E\cdot\nabla\Theta+ R.
\end{equation}
From \eqref{eq:const-rels}$_{1, 2}$ and assuming sufficiently smoothness on $\Psi$ with respect its arguments, 
\begin{equation}\label{eq:P-conj-eta}
\dfrac{\partial \eta}{\partial \mathbf{F}} = -\dfrac{\partial^2 \Psi}{\partial \mathbf{F}\partial \Theta}
=-\dfrac{\partial^2 \Psi}{\partial \Theta\partial \mathbf{F}}=-\dfrac{\partial \mathbf{P}}{\partial \Theta}.
\end{equation}
Using equation \eqref{eq:P-conj-eta} the conservation form \eqref{eq:cons-form} yields the field equation 
\begin{equation}
\rho_0 c_{_F} \dot{\Theta}=-\mathrm{Div}\mathbf{Q}+\dfrac{1}{\Theta}\mathbf{Q}_E\cdot\nabla\Theta + \Theta\dfrac{\partial \mathbf{P}}{\partial \Theta}:\dot{\mathbf{F}}+\rho_0 R,
\end{equation}
where $c_{_F}$ is the heat capacity and is defined by
\begin{equation}
\rho_0 c_{_F}=-\Theta\dfrac{\partial^2 \Psi}{\partial \Theta^2}.
\end{equation}
\section{The initial-boundary value problem}\label{sec:IBVP}
In this section we summarise the initial boundary-value problem for coupled non-classical thermoelasticity in a general non-linear framework.
\subsection{Local balance laws and constitutive equations}
The displacement--velocity relation, the balance of linear momentum \eqref{eq:Lmomu-bal}, the thermal displacement--temperature relation along with the balance of energy in the conservation form \eqref{eq:cons-form} lead to the first-order (in time) system of non-linear partial differential equation governing the strong coupling of thermoelastic behaviour in the non-classical regime, given by
\begin{equation}\label{eq:govern-eqn}
\left.
\begin{aligned}
\dot{\bm{u}} &=\bm{v}\\
\rho_0\dot{\bm{v}}&=\mathrm{Div}\mathbf{P}+\rho_0\mathbf{B}\\
\dot{\alpha}&=\Theta\\
\Theta\dot{\eta}&=-\mathrm{Div}\mathbf{Q}+\dfrac{1}{\Theta}\mathbf{Q}_E\cdot\nabla\Theta+ R
\end{aligned}
\right\rbrace\quad \text{ in }\quad \Omega\times\mathbb{I}.
\end{equation}
The constitutive equations are derived from a potential $\Psi$ denoting the classical Helmholtz free energy function, such that 
\begin{equation}\label{eq:summ-const}
\mathbf{P}=\dfrac{\partial \Psi}{\partial \mathbf{F}},\qquad \eta= -\dfrac{\partial \Psi}{\partial \Theta},\qquad\mathbf{Q}=\mathbf{Q}_E+\mathbf{Q}_D\quad\text{and}\quad\mathbf{Q}_E = -\Theta\dfrac{\partial \Psi}{\partial \bm{\Lambda}}.
\end{equation}
Equation \eqref{eq:summ-const} together with a constitutive relation for the dissipative component $\mathbf{Q}_D$ of  the heat flux satisfying an inequality of the form $\mathbf{Q}_D\cdot\nabla\Theta\leq 0$ complete the set of constitutive equations for system \eqref{eq:govern-eqn}.
\subsection{Initial and boundary conditions} 
Physically meaningful initial and boundary conditions in the non-classical theory are proposed in \cite{Wakeni2015}. To define initial conditions, we assume an initial configuration in which the thermal displacement field is homogeneous. 
Thus, we use initial conditions of the form 
\begin{equation}\label{eq:init-conds}
\bm{u}\vert_{_{t=0}}=\bm{u}^{0}, \quad \bm{v}\vert_{_{t=0}}=\bm{v}^{0}, \quad \alpha\vert_{_{t=0}}=0, \quad \text{ and }~ \Theta\vert_{_{t=0}}=\Theta^{0},
\end{equation}
where $(\,\cdot\,)^{0}$ are initially prescribed functions on $\Omega$.

For boundary conditions, we first consider the decomposition of the boundary $\Gamma$ into two mutually disjoint partition sets $\{\Gamma_{_{\bm{u}}},~\Gamma_{_\mathbf{T}}\}$, and $\{\Gamma_{_\Theta},~\Gamma_{_Q}\}$ such that
\begin{equation*}
\Gamma_{_{\bm{u}}}\cap~\Gamma_{_\mathbf{T}}=\Gamma_{_\Theta}\cap~\Gamma_{_Q}=\emptyset, ~~\text{and}~~\overline{\Gamma_{_{\bm{u}}}\cup~\Gamma_{_\mathbf{T}}}=
\overline{\Gamma_{_\Theta}\cup~\Gamma_{_Q}}\Gamma.
\end{equation*}
Hence physically meaningful boundary conditions have the form
\begin{equation}\label{eq:bound-conds}
\left.
\begin{aligned}
\bm{u}&=\check{\bm{u}}\quad \text{on }~\Gamma_{_{\bm{u}}}\times\mathbb{I},\\
\Theta &= \check{\Theta}\quad \text{on }~\Gamma_{_\Theta}\times\mathbb{I},
\end{aligned}\quad
\begin{aligned}
\text{and }~~\mathbf{P}\mathbf{N}&=\check{\mathbf{T}}\quad \text{on }~\Gamma_{_\mathbf{T}}\times\mathbb{I},\\
\text{and}~~\mathbf{Q}\cdot\mathbf{N}&=\check{Q}\quad \text{on }~\Gamma_{_Q}\times\mathbb{I},
\end{aligned}
\right.
\end{equation}
where $\check{{(\,\cdot\,)}}$ denote prescribed functions on the corresponding partitions of $\Gamma$.

Therefore, the system \eqref{eq:govern-eqn} together with the constitutive relations \eqref{eq:summ-const}, the initial and boundary conditions \eqref{eq:init-conds} and \eqref{eq:bound-conds} conclude the local description of the initial-boundary value problem (IBVP) for coupled non-classical thermoelasticity.
\section{Stability of the continuous problem. The Lyapunov function}\label{sec:stability}
Assume that there are no heat sources ($R=0$) and the part $\Gamma_{_{\Theta}}$ of the boundary  is fixed at a constant  
reference temperature $\Theta_0$ -- the temperature of the medium in which the motion of the solid takes place. Moreover, a thermally homogeneous boundary condition ($\mathbf{Q}\cdot\mathbf{N}=\check{Q}=0$) is prescribed on $\Gamma_{_Q}$, and the body and boundary mechanical loadings are conservative; that is, denoting the potential energy of the external loadings by $\Pi_{\mathrm{ext}}$, we have
\begin{equation}\label{eq:cons-loads}
\dfrac{\mathrm{d}}{\mathrm{d}t}\Pi_{\mathrm{ext}}=-\int_{\Omega}\rho_0\mathbf{B}\cdot\bm{v}~\mathrm{d}\Omega -\int_{\Gamma}\check{\mathbf{T}}\cdot\bm{v}~\mathrm{d}\Gamma.
\end{equation} 
Defining the relative temperature by $\vartheta=\Theta-\Theta_0$, the functional for the dynamics
\begin{equation}\label{eq:lyapunov}
\mathbf{\mathbb{E}}=\int_{\Omega}\big[\Psi+\vartheta\eta+\dfrac{1}{2}\rho_0\bm{v}\cdot\bm{v}\big]~\mathrm{d}\Omega +\Pi_{\mathrm{ext}},
\end{equation}
corresponding to the specific conditions just described above defines a Lyapunov function for the IBVP \eqref{eq:govern-eqn}. This can be shown by differentiating $\mathbb{E}$ with respect to time as
\begin{align}
\nonumber\dfrac{\mathrm{d}}{\mathrm{d}t}\mathbb{E} &= \int_{\Omega}\bigg[\mathbf{P}:\dot{\mathbf{F}}-\eta\dot{\Theta}-\dfrac{1}{\Theta}\mathbf{Q}_E\cdot\nabla\Theta+\dot{\vartheta}\eta+\vartheta\dot{\eta}+\rho_0\dot{\bm{v}}\cdot
\bm{v}\bigg]~\mathrm{d}\Omega + \dfrac{\mathrm{d}}{\mathrm{d}t}\Pi_{\mathrm{ext}}\\
\nonumber&=\int_{\Omega}\dfrac{\Theta_0}{\Theta^2}\mathbf{Q}_D\cdot\nabla\Theta~\mathrm{d}\Omega\\
&\leq 0. \label{eq:dLyapunov}
\end{align}
In the calculation leading to \eqref{eq:dLyapunov} we make use of the following: the time derivative of the free energy \eqref{eq:dfree2}, the linear momentum balance  \eqref{eq:Lmomu-bal} and \eqref{eq:cons-loads} together with integration by parts, and then the energy balance in conservation form \eqref{eq:cons-form} with integration by parts, and finally flux partitions and the constitutive restriction \eqref{eq:const-rels}$_{5}$ on the dissipative component $\mathbf{Q}_D$ of the heat flux.

Equation \eqref{eq:dLyapunov} implies that the positive definite functional $\mathbb{E}$ is monotonically decreasing along the flow defined by the IBVP, which proves the non-linear stability of the continuous problem for the specific case outlined above. A similar result for the case of more general boundary conditions is not known, even in the classical case (see also \cite{Armero1992}). 
\section{The linearized theory}\label{sec:linear-theory}
The IBVP \eqref{eq:govern-eqn} is linearized about the reference configuration for which $\bm{u}=\mathbf{0}$, $\bm{v}=\mathbf{0}$, $\alpha = 0$, and $\Theta=\Theta_0$ are assumed to be natural for the configuration. As a consequence, the linearized version of IBVP \eqref{eq:govern-eqn} is given by
\begin{equation}\label{eq:lin-gov-eqn}
\begin{aligned}
\dot{\bm{u}} &= \bm{v},\\
\rho\bm{v} &= \mathrm{div}\bm{\sigma}+\rho\bm{b},\\
\dot{\alpha} &= \Theta,\\
\Theta_0\dot{\eta}&= -\mathrm{div}\bm{q}+r,
\end{aligned}
\end{equation} 
with constitutive equations
\begin{equation}\label{eq:lin-cons-rels}
\bm{\sigma} = \dfrac{\partial \psi}{\partial \bm{\pmb{\varepsilon}}}, \quad\eta = -\dfrac{\partial \psi}{\partial \Theta},\quad \bm{q} = \bm{q}_e + \bm{q}_d, \quad \bm{q}_e = -\Theta_0\dfrac{\partial \psi}{\partial \Lambda},~~\text{ and }~\bm{q}_d = -\mathbf{K}_2\nabla\Theta, 
\end{equation}
where $\psi$ is the quadratic free energy function defined by
\begin{equation}\label{eq:lin-free-ener}
\psi = \dfrac{1}{2}\bm{\pmb{\varepsilon}}:\mathbb{C}\bm{\pmb{\varepsilon}}-\vartheta\mathbf{m}:\bm{\pmb{\varepsilon}}-\dfrac{\rho c}{2\Theta_0}\vartheta^2+\dfrac{1}{2}\mathbf{K}_1\nabla\alpha\cdot\nabla\alpha.
\end{equation}
Then, by \eqref{eq:lin-cons-rels} and \eqref{eq:lin-free-ener}, we obtain
\begin{equation}
\bm{\sigma} = \mathbb{C}\bm{\pmb{\varepsilon}}-\vartheta\mathbf{m},\quad \eta =\dfrac{\rho c}{\Theta_0}\vartheta+\mathbf{m}:\bm{\pmb{\varepsilon}},~~\text{and }~\bm{q}=-\Theta_0\mathbf{K}_1\nabla\alpha-\mathbf{K}_2\nabla\Theta. 
\end{equation}
Here, $\bm{\sigma}$, $\bm{q}$, $\bm{\pmb{\varepsilon}}$, $\mathbf{m}$ are the stress tensor, the heat flux, the infinitesimal (symmetric) strain, and thermomechanical coupling tensor, respectively. Under the assumption of isotropy, the fourth-order elasticity tensor $\mathbb{C}$, and the non-classical conductivity tensor $\mathbf{K}_1$ are both symmetric and positive-definite, while the classical heat conduction tensor $\mathbf{K}_2$ is symmetric and positive-semidefinite.

Note the similarity between this linearized model and the type III model by Green and Naghdi (see for example \cite{Wakeni2015}). The only difference is that the non-classical conductivity tensor $\mathbf{K}_1$ is replaced by $\Theta_0\mathbf{K}_1$. However, the model due to Green and Naghdi is not consistent with the laws of thermodynamics in the non-linear case.   
\section{Conclusion}
A thermodynamically consistent formulation for coupled thermoelasticity at finite strain in a generalized framework was developed. The model encompasses the classical model based on Fourier's law of heat conduction as well as the non-classical one in which a non-Fourier type heat conduction equation allows for the propagation of thermal energy as a wave (with finite speed). The formulation is based on two key postulates: the first is that the introduction of a scalar internal state variable, the thermal displacement, defined as the time primitive of the absolute temperature, and the second is that the heat flux is composed of both energetic and dissipative components. Physically meaningful initial and boundary conditions were presented. A Lyapunov type functional for the governing IBVP was suggested for some class of boundary conditions. It was shown that the system is non-linearly stable. The linearized model was shown to be similar to the type III model of Green and Naghdi.

The design and analysis of efficient numerical algorithms for the non-linear thermomechanical model will be described in future work. For such non-linear and strongly coupled problem, it seems natural to use of shock capturing schemes such as the space-time discontinuous Galerkin approach together with an operator-splitting technique to effectively decouple the system of governing equations without violating the stability of the full system.      

{\bf Acknowledgments.} The work reported in this paper has been supported by the National Research
Foundation of South Africa through the South African Research Chair in Computational Mechanics. This
support is acknowledged with thanks.
\section*{References}


\end{document}